\documentclass[conference]{IEEEtran}
\IEEEoverridecommandlockouts
\usepackage{cite}
\usepackage{amsmath,amssymb,amsfonts}
\usepackage{amsthm}
\usepackage{algorithmic}
\usepackage{graphicx}
\usepackage{textcomp}
\usepackage{xcolor}
\usepackage{framed}
\usepackage{fancyvrb}
\usepackage{fvextra}
\usepackage{multirow}
\usepackage[colorlinks=true, linkcolor=blue, urlcolor=blue, citecolor=blue]{hyperref}
\usepackage{xurl}
\def\BibTeX{{\rm B\kern-.05em{\sc i\kern-.025em b}\kern-.08em
    T\kern-.1667em\lower.7ex\hbox{E}\kern-.125emX}}

\newtheorem{hypothesis}{\textbf{Hypothesis}}

\begin{document}

% \title{Improving Trust in Government Services: A Comparative Analysis of LLM-Generated Tax Refund Explanations*
% \thanks{\textsuperscript{*}Identify applicable funding agency here. If none, delete this.}
% }

%Selecting the Right LLM for E-Government Service Explanations
%GenAI for e-Government: Comparing LLM Tax Refund Explanations
%Leveraging Generative AI in e-Government Services: A Comparative Analysis of LLM-Generated Tax Refund Explanations
%GenAI for e-Government: A Comparative Analysis of LLM Tax Refund Explanations
% How to select an LLM for e-government services explanations
%Leveraging LLMs for e-government services explanations
%GenAI for e-Government: A Comparative Analysis of LLM Tax Refund Explanations

\title{Selecting the Right LLM for eGov Explanations\thanks{This project has received funding from the European Union’s Horizon research and innovation programme under grant agreements no 101094905 (AI4GOV) and 101092639 (FAME).}
}

% \author{
% \centering
% \IEEEauthorblockN{Fabiana Fournier}
% \IEEEauthorblockA{\textit{IBM Research} \\
% Israel \\
% fabiana@il.ibm.com}
% \and
% \IEEEauthorblockN{Lior Limonad}
% \IEEEauthorblockA{\textit{IBM Research} \\
% Israel \\
% liorli@il.ibm.com}
% \and
% \IEEEauthorblockN{Hadar Mulian}
% \IEEEauthorblockA{\textit{IBM Research} \\
% Israel \\
% hadar.mulian@ibm.com}
% \and
% \and
% \and

\author{
\IEEEauthorblockN{Lior Limonad\IEEEauthorrefmark{1},
Fabiana Fournier\IEEEauthorrefmark{1},
Hadar Mulian\IEEEauthorrefmark{1},
George Manias\IEEEauthorrefmark{2},
Spiros Borotis\IEEEauthorrefmark{3} and
Danai Kyrkou\IEEEauthorrefmark{4}}

\IEEEauthorblockA{\IEEEauthorrefmark{1}IBM Research, Israel \\ Emails: liorli@il.ibm.com, fabiana@il.ibm.com, hadar.mulian@ibm.com}
\IEEEauthorblockA{\IEEEauthorrefmark{2}Department of Digital Systems,
University of Piraeus, Piraeus, Greece\\
Email: gmanias@unipi.gr}
\IEEEauthorblockA{\IEEEauthorrefmark{3}Maggioli S.p.A., Santarcangelo di Romagna, Italy\\
Email: spiros.borotis@maggioli.gr}
\IEEEauthorblockA{\IEEEauthorrefmark{4}ViLabs, Limassol, Cyprus\\
Email: danaikyrkou@vilabs.eu}
}

\maketitle

\maketitle

%%%%%%%%%%%%%%%%%%%%%%%%%%%%%%% Comment management
\newif\ifshowcomments
\showcommentstrue
%\showcommentsfalse
\ifshowcomments
    \newcommand{\mynote}[2]{\fbox{\bfseries\sffamily\scriptsize{#1}}{\small$\blacktriangleright$\textsf{#2}$\blacktriangleleft$}}
\else
    \newcommand{\mynote}[2]{}
\fi
\newcommand{\is}[1]{\textcolor{blue}{\mynote{Inna}{#1}}}
\newcommand{\lior}[1]{\textcolor{blue}{\mynote{LIOR}{#1}}}
\newcommand{\ff}[1]{\textcolor{purple}{\mynote{Fabiana}{#1}}}
\newcommand{\yd}[1]{\textcolor{orange}{\mynote{Yuval}{#1}}}
%%%%%%%%%%%%%%%%%%%%%%%%%%%%%%%%

\begin{abstract}
The perceived quality of the explanations accompanying e-government services is key to gaining trust in these institutions, consequently amplifying further usage of these services. Recent advances in generative AI, and concretely in Large Language Models (LLMs) allow the automation of such content articulations, eliciting explanations' interpretability and fidelity, and more generally, adapting content to various audiences. However, selecting the right LLM type for this has become a non-trivial task for e-government service providers. In this work, we adapted a previously developed scale to assist with this selection, providing a systematic approach for the comparative analysis of the perceived quality of explanations generated by various LLMs. We further demonstrated its applicability through the tax-return process, using it as an exemplar use case that could benefit from employing an LLM to generate explanations about tax refund decisions. This was attained through a user study with 128 survey respondents who were asked to rate different versions of LLM-generated explanations about tax refund decisions, providing a methodological basis for selecting the most appropriate LLM. Recognizing the practical challenges of conducting such a survey, we also began exploring the automation of this process by attempting to replicate human feedback using a selection of cutting-edge predictive techniques.
\end{abstract}


\section{Introduction}

The concept of e-government refers to using Information and Communication Technologies (ICTs) to deliver government services to citizens and businesses more effectively and efficiently. It is the application of ICT in government operations, achieving public ends by digital means\footnote{\url{https://publicadministration.un.org/egovkb/en-us/Overview\#whatis}}. It enables access by citizens, business units, governments, and others to services online without the involvement of a third party~\cite{Alminshid2021FactorsSector}. 

Including innovative e-government services also brings about the need to build citizen trust in their use. Lack of trust has long been recognized as an impediment to the adoption of e-government services. Furthermore, the level of trust citizens have in e-government has been seen as a key barrier to greater usage of online government services~\cite{Dashti2010TrustPerspective,Carter2005TheFactors}.
Moreover, the authors in~\cite{Dashti2010TrustPerspective} state that trust in e-government is not solely a function of trust in technology and government but is also influenced by the perceived responsibility of e-government and users’ perceptions of the trust conveyed through its design features, functions, and processes.

AI-Augmented Business Process Management Systems (ABPMSs)~\cite{Dumas2023} represent a new generation of business process management tools designed to enhance process execution with AI-driven capabilities. However, ``AI can be developed and adopted only if it satisfies the stakeholders' and users' expectations and needs, and that is how the role of trust becomes essential''~\cite{Afroogh2024}. A key feature of ABPMSs is their inherent trustworthiness, supported by their ability to explain and reason about process executions. However, providing accurate explanations is challenging because it requires the ability to reflect the evolving conditions and the context in which decisions were made during process execution. Often, explanations depend on reasoning about broader situational factors, not just current tasks or decisions. In ABPMSs, explainability is a key characteristic realized via Situation-Aware eXplainability (SAX). SAX focuses on the need to provide sound and interpretable explanations regarding process executions. In particular, recent advances in AI present a new opportunity to leverage a family of generative AI models, known as Large Language Models (LLMs), to facilitate the automatic generation of such explanations.

As for e-government systems, trust is an essential component in the adoption of ABPMSs and their successful integration into the operational systems of organizations and institutions. In this context, explainability can be seen as a key component of trust and as the glue connecting the strategies driven by organizational systems with users' intentions to adopt and adhere to the regulations underlying the operation of such systems. Explanations can also serve to either expose and sometimes mitigate the consequences of applying different policies and changes in a strategy. Consequently, trustworthiness is formed by the perceived quality of the explanations produced by the systems. 

At the heart of SAX is the perception that explainability in business processes (BPs) promotes trust and adoption of automation technology. SAX seeks explanations of business process outcomes and conditions that are sound and interpretable taking advantage of the BP definitions, contextual environment, and full runtime process traces. Furthermore, they are expected to embed the ability to go beyond a local reasoning context, handle a large variety of situations, and facilitate the automatic tracking of execution consistency for a better understanding of process flows and process outcomes. One way to provide automatic explanations is by exploiting LLMs' capabilities.

When dealing with e-government systems, ``...ignoring users can lead to poorly designed systems, resistance to change, and even outright refusal, which ultimately results in a failure for the users.''~\cite{Abdul2024}.
In this work, we address the issue of users' trust in e-government processes by involving users' perceptions of the quality of explanations issued to them automatically by the ABPMS. To this end, we leverage our previous work~\cite{Fahland2024} 
in which we developed and validated a novel scale for pragmatic assessment of the quality of explanations as perceived by users of business processes. Our study explored the quality of business process explanations in natural language, testing whether the users perceive LLM-generated text as being clear and sound. A key innovation in our study was using causal execution knowledge to generate better explanations~\cite{Fournier2023v3}.

This work tackles the challenge of selecting an LLM to facilitate explainability automation in e-government systems.
We address this problem by leveraging our previous research by adopting and applying the developed scale to the process of tax refund as an exemplar of an e-government system. We demonstrate that using our scale provides a quantitative means to compare alternative LLMs, enabling users to select the one perceived as the best. Realizing that engagement with citizens is a significant burden, we extended this contribution by exploring the viability of automating the acquisition of ``crowd opinion'' by leveraging AI techniques as an alternative.

As LLMs will become prevalent in everyday life in the coming years, the question of which LLM to apply will become more relevant. In fact,
Market research indicates significant growth in the LLM sector. Grand View Research estimates the global LLM market size at USD 4.35 billion in 2023, with a projected compound annual growth rate (CAGR) of 35.9\% from 2024 to 2030\footnote{\url{https://www.grandviewresearch.com/industry-analysis/large-language-model-llm-market-report}}.

\section{Background}

Artificial intelligence (AI) has various applications in government. It can help advance public policy goals in areas like emergency services, health, and welfare while also improving public interactions with the government through virtual assistants. One example of the latter is ``Skatti'', an AI-powered chatbot developed by the Swedish Tax Agency (Skatteverket) to enhance customer service by assisting citizens with tax-related inquiries\footnote{\url{https://www.ai.se/en/news/artificial-intelligence-improves-swedish-tax-agencys-customer-service}}. 

One way to benefit from AI in e-government services is through the automation of business processes. Business Process Management (BPM) is a managerial effort devoted to ensuring the constant, efficient, and effective movement of a business entity towards its constantly changing goals. This entails adopting a process view of the business and includes the planning, analysis, design, implementation, execution, and control of BPs. A business process (BP) is a collection of tasks that are executed in a specific sequence to achieve some business goal~\cite{Weske2019}. The digital footprint that depicts a single execution of a process as a concrete sequence of activities or events is termed a `trace'~\cite{vanderAalst2016ProcessMining}. A multi-set of traces is usually referred to as a trace-log or event-log.

% \ff{I don't think we can use the following text, just in case I am leaving it for now}
% In~\cite{Chinnasamy2023E-GovernenceTechniques} the authors mention confidence among the difficulties in adopting e-government applications. By confidence they mean ``believing in online services strongly depends on a few factors, such as people' trust in the government as a whole, the nature of the online
% services, and the individuals' personal beliefs (e.g., there still a large number of citizens who prefer to
% handle paper applications rather than web services).''

ABPMSs are a special case of BPM systems where the business process execution is entangled with the use of various AI technologies, and where such embedding makes the explainability of AI even more challenging as it should consider the process context in which AI operates. In addition, advancements in AI have led to the development of more sophisticated yet increasingly complex models. 

Explainable AI (XAI) seeks to address the opacity of these models by offering explanations for the decisions and outcomes generated by the AI system. State-of-the-art AIX frameworks are predominately developed for post-hoc interpretations of Machine Learning (ML) models~\cite{Adadi2018,Meske2022}. Context-wise, they can be divided into global, local, and hybrid explanations~\cite{Adadi2018,Rehse2019,Guidotti2018}. Global explanations attempt to explain the ML model’s internal logic, local explanations try to explain the ML model’s prediction for a single input instance, and hybrid approaches vary (e.g., explaining the ML model’s internal logic for a subspace of the input space).

Recent advancements in AI have opened new opportunities for implementing explainability using generative AI (GenAI) models. 
Generative AI refers to AI techniques that learn a representation of artifacts from data, and use it to generate brand-new, unique artifacts that resemble but don’t repeat the original data~\cite{GartnerGenAI}.
Specifically, in the field of Natural Language Processing (NLP), advanced AI models, known as LLMs, have been transformative in this regard. Leveraging LLMs for explainability eliminates the need for training dedicated explainability-focused ML models, significantly reducing the associated effort and complexity.

One way to automate explanation is to exploit Large Language Models' (LLMs) capabilities. As supported by a recent Gartner report~\cite{Gartner2024}, LLMs are being adopted across all industries and business functions, driving a variety of use cases such as text summarization, question-answering, document translation, and alike. LLMs can also be augmented by additional capabilities to create more robust systems and feature a growing ecosystem of tools. Large language models (LLMs) are a specialized type of GenAI models trained on extensive text data to handle various natural language processing tasks~\cite{GartnerLLM}. One of their key strengths is the ability to perform few-shot and zero-shot learning through prompt-based techniques~\cite{LLMprompt2023}. Recently, the field of ``prompt engineering'' has emerged, focusing on creating, refining, and applying instructions to guide LLM outputs effectively. This practice aims to optimize LLM utilization by tailoring it to users' skills and the context of tasks like analysis or reasoning. Techniques in prompt engineering include using one-shot or few-shot examples, annotating prompts with quotes, employing methods like ``Chain of Thought'' (CoT), ``LLM-as-a-Judge'', Retrieval Augmented Generation (RAG), and adjusting LLM settings such as temperature and top-p sampling to control variability.

SAX aims to generate explanations about BPs, considering multiple knowledge perspectives. This includes knowledge about the process model, process execution traces, causal execution dependencies~\cite{Fournier2023v3}, and XAI attribute ranking. Leveraging the power of LLMs, these elements are derived and synthesized to generate process and context-aware explanations, namely SAX explanations. SAX services are realized in the SAX4BPM open-source library available at: \url{https://github.com/IBM/Sax4bpm}.

\section{The TAX Refund Case}

\begin{figure*}[htbp]
\centerline{\includegraphics[width=0.8\textwidth]{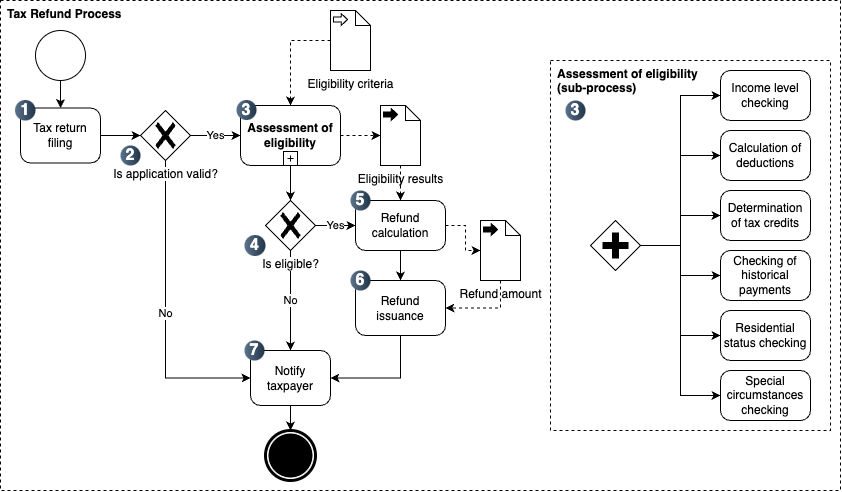}}
    \vspace{-1em}
\caption{Tax refund process.}
    \vspace{-1.5em}
\label{fig:tax-refund-process}
\end{figure*}

We use the tax refund process as a representative case for a business process that facilitates services issued by the government to the citizens (i.e., taxpayers). It is a relatively known and mandatory process commonly used across countries globally. The main steps, illustrated in Figure~\ref{fig:tax-refund-process}, are general enough so they can be understood by users worldwide.

The tax refund process (see Figure~\ref{fig:tax-refund-process}) begins when the taxpayer files their tax return, submitting necessary financial documents such as income statements and receipts for deductions (step 1). Once the return is filed, the tax authority validates it (step 2) by reviewing its completeness, accuracy, and checking for missing or invalid information. If the file is not valid, a notification is sent to the taxpayer (step 7). If it is valid, the tax authority proceeds to assess the taxpayer’s refund eligibility (step 3) based on factors like income level, eligible deductions (e.g., mortgage interest, medical expenses), tax credits, tax payments made, residency status, and any exceptional circumstances. If the taxpayer qualifies for a refund (step 4), the tax authority calculates the refund amount (step 5), considering overpaid taxes, deductions, credits, residency adjustments, and any special circumstances. The final refund amount is then issued (step 6), and the taxpayer is notified of their refund status (step 7). The process ends once the refund is issued or if the taxpayer is required to meet additional eligibility requirements. A detailed version of this process was articulated as a nested textual form and included in the LLM prompt (see Section~\ref{sec:instrumentation}).

Tax refund filing may ultimately be rejected for a large variety of reasons as implied from the illustrated process. For practical reasons, we considered three specific conditions under which an applicant might further inquire about their application. The first condition, ``correct and resubmit'', arises when the file lacks income statements for March and June during the requested income return period. The second, ``\$100 short'', pertains to situations where a historical debt of \$100 was identified, which the applicant may have been unaware of. The third, ``why is it not progressing'', involves a possible inquiry regarding an application still under processing (``in-flight'') where the decision appears to be delayed.

\section{Hypothesis}

Our intention is to systematically address the challenge of selecting the most appropriate LLM among multiple LLMs to automatically generate explanations concerning decisions made by a tax refund system. Most appropriate here means an LLM that automates the articulation of trustworthy explanations that promote usage.

To test this hypothesis, we have set up our empirical effort to demonstrate that citizens' perceptions of an LLM's explanation quality can be captured and subsequently used as a proxy to practically guide the choice among competing LLMs in an e-government IT infrastructure that employs GenAI for explainability. More concretely, we formulated our research hypothesis as follows:

\begin{hypothesis}
The perceived quality of LLM-generated explanations can be empirically elicited to determine the ranking among a set of alternative LLM model types.
\end{hypothesis} 

Using our previously developed scale~\cite{Fahland2024}, 
we elicited concrete perceptions of \textit{fidelity} and \textit{interpretability} through self-reporting. This allowed us to compare a set of LLM model types and determine which performs best as perceived by the users. In line with the same scale, we also inquired about participants' \textit{trust} in LLMs and their overall \textit{curiosity} about receiving an explanation for the presented cases—two potential moderating factors that may influence the main effect on the perceived quality of the explanations. Given its widespread distribution in our sample, %(see Figure~\ref{fig:bpm-expertise}), 
we extended the analysis to include \textit{BPM expertise} as an additional covariate.

\section{Method}

We conducted a controlled experiment, employing an online developed survey accessible via Eusurvey\footnote{\url{https://eusurvey.escoaladevalori.ro/eusurvey/home/welcome}}, to elicit user ratings concerning their perception of the quality of a variety of textual narratives automatically generated by LLMs in response to different inquiries about the aforementioned tax refund process conditions. These ratings were then carefully inspected to facilitate a comparative analysis among the explanations, serving as the proxy for choosing the best LLM model to be employed in a system that can automate the generation of such explanations. The survey forms, prompts, and collected data are accessible at~\cite{Survey2025}.

\subsection{Participants}

We began recruiting individuals through the AI4GOV\footnote{\url{https://ai4gov-project.eu/}} EU project, initially involving project partners and pilot participants, and later expanded outreach through their networks. While this approach may not constitute random sampling, we ensured the sample's representativeness of the general taxpayer population, as evidenced by 97.7$\%$ of participants who identified themselves as taxpayers.
% We targeted the population of the various individuals participating in the AI4GOV\footnote{\url{https://ai4gov-project.eu/}} EU project, including project partners and pilots. The majority of this group faithfully represents the general population of taxpayers, as indicated by 97.7$\%$ of the participants who responded positively to whether they identify themselves as taxpayers. 
Participation was facilitated through email invitations disseminated by the project coordinator and lead partner representatives, who ensured a relatively balanced random allocation to the different groups. The email invitations and the text displayed at the outset of the online survey clearly stated that participation was voluntary and anonymous, with no personal information being collected. Before participating, users also provided their consent. As this study involved anonymized survey responses and no direct participant intervention, ethics approval was not required according to institutional guidelines.

The survey included a total of 128 participants. Regarding gender, 76 were men (59.4\%), 50 were women (39.1\%), and 2 chose not to specify their gender (1.6\%). We also inquired about the level of education, and our sample primarily consisted of individuals with graduate (76.6\%) and undergraduate (18.8\%) education. On a 1-7 scale, most participants rated their digital literacy level as 5 or higher (93\%). Regarding general business process management background, distribution was relatively widespread.% as illustrated in Figure~\ref{fig:bpm-expertise}.

% \begin{figure}[htbp]
% \centerline{\includegraphics[width=\columnwidth]{figures/bpm-expertise.png}}
% \vspace{-1em}
% \caption{BPM Expertise.}
% \vspace{-1em}
% \label{fig:bpm-expertise}
% \end{figure}

\subsection{Experimental Design}

% We conducted a mixed design 4X3 controlled experiment to assess our hypothesis. This included a between-group manipulation of four LLM model types used to populate explanations for three different case inquiries about the tax refund process. Each participant was assigned to respond on two different explanations populated by two different LLM models. 
We conducted a mixed-design 4×3 controlled experiment to test our hypothesis. The study involved a between-subjects manipulation of four LLM model types used to generate explanations for three different case inquiries related to the tax refund process. Each participant was assigned to evaluate two different explanations, each generated by a different LLM model. We employed and adapted the scale previously developed in~\cite{Fahland2024} 
to capture participants' perceptions of each explanation. The scale comprises six underlying self-reported ratings: completeness, soundness, causability, clarity, compactness, and comprehensibility. The former three form the perception of fidelity, while the latter three form the perception of interpretability. The scale was also originally extended to include measurements of trust and curiosity, two covariates that have been found to influence the perceived quality of explanations. The survey consists of 24 statements, each rated on a 1-7 Likert scale. For this study, the wording of the statements was adapted to match the tax refund process. For example, the original statement: ``The explanation encompasses all aspects and outcomes related to the condition.'' was rephrased to ``The explanation encompasses all aspects related to the status of your application.'' 

\subsection{Instrumentation}
\label{sec:instrumentation}

Overall, the 24 statements mentioned above were embedded in an anonymous online survey, which included an overview section describing the survey, a participation consent, a set of background-related questions, a description of the tax refund process, and two inquiry cases. Background questions included gender, education level, expertise in business processes, and digital literacy. 
Each case presented a query about a specific process condition, accompanied by a textual narrative generated by an LLM to explain it and a corresponding `ground truth' version for the same case, articulated by the researchers. Participants were instructed to read each explanation and evaluate its quality relative to the `ground truth' version across a series of statements.

Four contemporary LLM models were employed to generate the explanations: granite-3-8b-instruct (Apache 2.0), llama-3-1-70b (MIT), flan-ul2-20b (Apache 2.0), and GPT-4o (OpenAI). %As noted in brackets, the first three are open-source models available on the Hugging Face platform, whereas the latter has a proprietary model offered for free but with usage limits. 
We selected these LLMs to represent a diverse cross-section of the current LLM landscape in terms of scale (ranging from 8B to 70B parameters), architecture (e.g., encoder-decoder in FLAN-UL2 vs. decoder-only in LLaMA 3), licensing (open-source models like Granite and FLAN vs. proprietary GPT-4o), and origin (community-driven vs. big tech labs). This selection enables meaningful comparisons that reflect the breadth of LLM design and deployment strategies.

Designated LLM prompting was employed to generate the explanations. Each prompt followed the template outlined in Figure~\ref{lst:prompt}. The template included a description of the tax refund process, an enumeration of the activities within the process, a description of the causal dependencies among these activities, and an instruction to generate an explanation related to a specific decision or condition of the tax refund process. For the prompt addressing the ``why is it not progressing'' case, since this scenario pertains to an `in-flight' process, it also incorporated the trace log of tasks already executed, along with their timestamps, preceding the instructional entry. 

% Define a custom verbatim environment with fancyvrb
\DefineVerbatimEnvironment{MyVerbatim}{Verbatim}{commandchars=\\\{\}, breaklines=true, breaksymbol={}, breakindent=0pt}

\begin{figure}[ht] % Use H to enforce placement here
\begin{framed}
\begin{MyVerbatim}
\textbf{PROCESS DESCRIPTION:} The following is a description of a national task refund process ...
\textbf{Process Model:} 1. Start: Taxpayer Files Tax Return. ... 2. Validation of Tax Return. ... 3. ...
\textbf{Causal Model:} (tax return filing > validation of application) and ... [where '>' means causes] 
\textbf{INSTRUCTION:} You are appointed the role of generating explanations in response to individual applicants' inquiries about their tax refund decisions.
With respect to the above process given a particular applicant with the following profile: Name: ...
\textbf{The decision was:} Notify taxpayer to correct and resubmit. 
\textbf{Generate an explanation that replies to the following query:} Why was I notified to correct my form and resubmit?
The explanation should not exceed 150 words. Do not include any required actions.
\end{MyVerbatim}
\end{framed}
\vspace{-1em}
\caption{Prompt example for explanation generation}
\vspace{-1em}
\label{lst:prompt}
\end{figure}

\section{Results}

\begin{figure}[htbp]
\centerline{\includegraphics[width=\columnwidth]{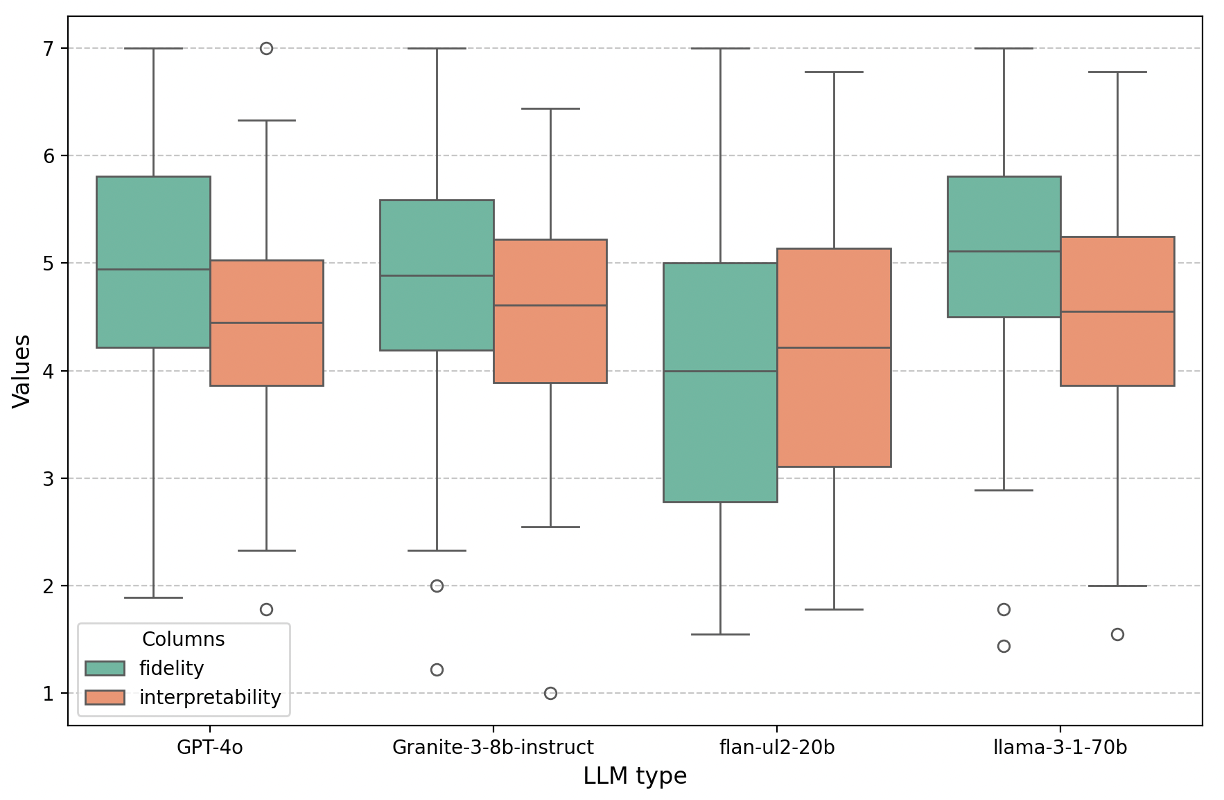}}
    \vspace{-1em}
\caption{Box-plot distributions per LLM model type}
    \vspace{-1em}
\label{fig:models-boxplots}
\end{figure}

Descriptive statistics for all LLMs corresponding to explanation quality as perceived by fidelity and interpretability are detailed in Table~\ref{tab:descriptives}. Box-plot analysis of LLM-wise distributions of fidelity and interoperability is shown in Figure~\ref{fig:models-boxplots}. As illustrated, for three out of the four models, fidelity scores were higher than interoperability scores. However, this trend was reversed for the Flan model, likely due to its explanations being significantly more compact across all three cases.

\begin{figure}[htbp]
\centerline{\includegraphics[width=\columnwidth]{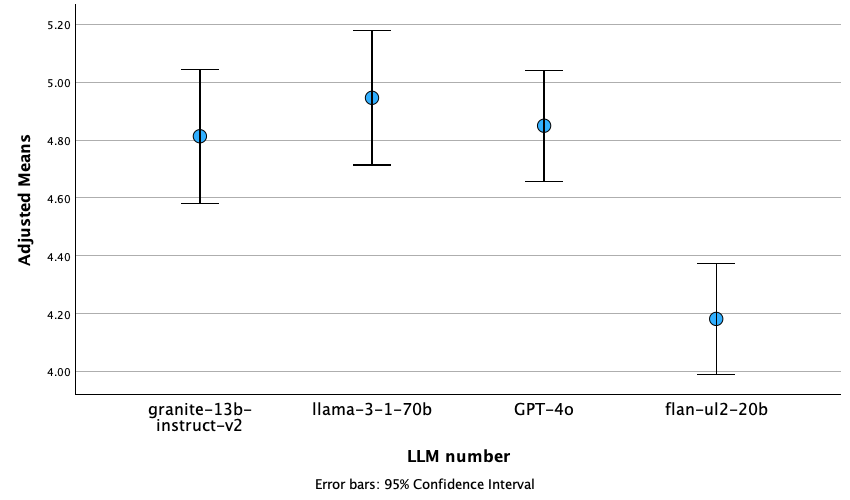}}
    \vspace{-1em}
\caption{Adjusted Means for Fidelity}
    \vspace{-1em}
\label{fig:adjusted-means-fidelity}
\end{figure}

\begin{figure}[htbp]
\centerline{\includegraphics[width=\columnwidth]{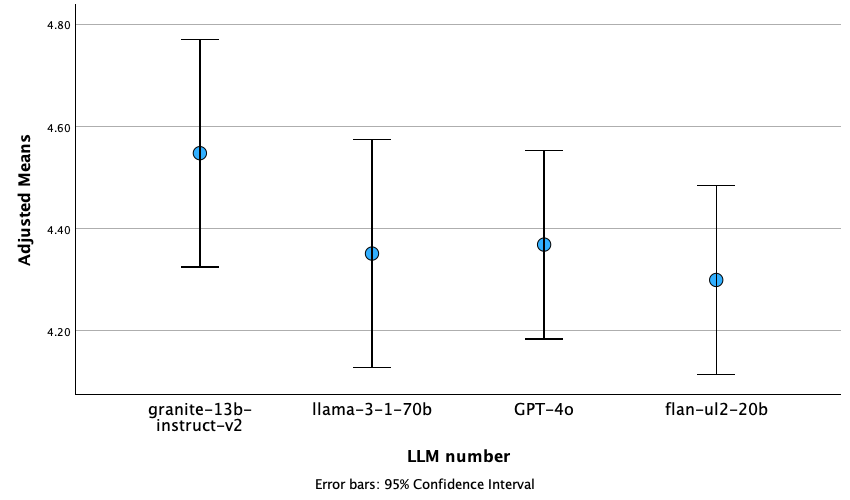}}
    \vspace{-1em}
\caption{Adjusted Means for Interpretability}
    \vspace{-1.5em}
\label{fig:adjusted-means-interpretability}
\end{figure}

Results for the analysis of the survey are detailed in Table~\ref{tab:mancova}. In line with the original scale development, we included trust and curiosity as covariates in the analysis. To this, we also added digital literacy and BPM competence as two additional measurements that were acquired in our current survey. An analysis of covariance (MANCOVA) was conducted to assess the effects of LLM-type manipulation on the two dependent variables, fidelity and interpretability, while controlling for the four covariates. The analysis showed that for both dependent variables, the effect of the manipulation was statistically significant. However, examination of the covariates showed, consistent with the original scale development work, that only trust and curiosity had a significant effect as covariates, whereas digital literacy and BPM competence did not have a significant effect as covariates. Thus, with the inclusion of trust and curiosity, the differences among the LLMs were manifested by the mean differences in the perceptions about the quality of the explanations. Adjusting for the covariates, these differences are as illustrated in Figures~\ref{fig:adjusted-means-fidelity} and~\ref{fig:adjusted-means-interpretability}.

\begin{table}[htbp]
\vspace{-1.1em}
\caption{Descriptive statistics for all LLM types}
\label{tab:descriptives}
%\resizebox{\columnwidth}{!}{%
\begin{tabular}{|l|l|l|l|l|}
\hline
 & \textbf{LLM type} & \textbf{Mean} & \textbf{SD} & \textbf{n} \\ \hline
\textbf{Fidelity} & (1) granite-3-8b-instruct & 4.8073 & 1.174 & 52 \\ \hline
 & (2) llama-3-1-70b & 5.0473 & 1.14744 & 52 \\ \hline
 & (3) GPT-4o & 4.9309 & 1.14432 & 76 \\ \hline
 & (4) flan-ul2-20b & 4.035 & 1.38178 & 76 \\ \hline
 & Total & 4.6635 & 1.28801 & 256 \\ \hline
\textbf{Interpretability} & (1) granite-3-8b-instruct & 4.5408 & 0.99887 & 52 \\ \hline
 & (2) llama-3-1-70b & 4.4419 & 1.12014 & 52 \\ \hline
 & (3) GPT-4o & 4.4443 & 1.05138 & 76 \\ \hline
 & (4) flan-ul2-20b & 4.1664 & 1.33939 & 76 \\ \hline
 & Total & 4.3809 & 1.1507 & 256 \\ \hline
\end{tabular}%
%}
\end{table}

As observed in these figures, the charts can help determine which LLM type performs best. However, considering perceived fidelity, the differences among the top three models seem somewhat insignificant, particularly when compared to the Flan model. Similarly, for perceived interpretability, the Granite model scores the highest, while the differences among the other three models also appear negligible. Subsequent pairwise comparisons using contrast analysis corroborated these observations, revealing significant differences only between the best and worst models in the case of fidelity (p $<$ .001) and a mildly significant difference between the best and worst models in the case of interpretability (p $<$ .1).

\begin{table*}[htbp]
\centering
\caption{ANCOVA (i.e., including curiosity*, trust*, digital literacy, and BPM competence as covariates) results for the tax refund case. When controlling for curiosity and trust, both effects on fidelity and interpretability were
significant.}
\label{tab:mancova}
%\resizebox{\textwidth}{!}{%
\begin{tabular}{|l|llllllllllll|}
\hline
\textbf{} & \multicolumn{12}{l|}{\textbf{MANCOVA}} \\ \hline
\textbf{} & \multicolumn{2}{l|}{\textbf{Granite (1)}} & \multicolumn{2}{l|}{\textbf{llama (2)}} & \multicolumn{2}{l|}{\textbf{GPT-4o (3)}} & \multicolumn{2}{l|}{\textbf{flan (4)}} & \multicolumn{1}{l|}{\multirow{2}{*}{\textbf{df}}} & \multicolumn{1}{l|}{\multirow{2}{*}{\textbf{F}}} & \multicolumn{1}{l|}{\multirow{2}{*}{\textbf{Sig.}}} & \multirow{2}{*}{\textbf{$\eta^2$}} \\ \cline{1-9}
\textbf{Constructs} & \multicolumn{1}{l|}{\textbf{M}} & \multicolumn{1}{l|}{\textbf{SE}} & \multicolumn{1}{l|}{\textbf{M}} & \multicolumn{1}{l|}{\textbf{SE}} & \multicolumn{1}{l|}{\textbf{M}} & \multicolumn{1}{l|}{\textbf{SE}} & \multicolumn{1}{l|}{\textbf{M}} & \multicolumn{1}{l|}{\textbf{SE}} & \multicolumn{1}{l|}{} & \multicolumn{1}{l|}{} & \multicolumn{1}{l|}{} &  \\ \hline
Fidelity & \multicolumn{1}{l|}{4.81} & \multicolumn{1}{l|}{0.12} & \multicolumn{1}{l|}{4.95} & \multicolumn{1}{l|}{0.12} & \multicolumn{1}{l|}{4.85} & \multicolumn{1}{l|}{0.1} & \multicolumn{1}{l|}{4.18} & \multicolumn{1}{l|}{0.1} & \multicolumn{1}{l|}{(7,248)} & \multicolumn{1}{l|}{48.41} & \multicolumn{1}{l|}{*\textless{}.001} & 0.577 \\ \hline
Interpretability & \multicolumn{1}{l|}{4.55} & \multicolumn{1}{l|}{0.11} & \multicolumn{1}{l|}{4.35} & \multicolumn{1}{l|}{0.11} & \multicolumn{1}{l|}{4.37} & \multicolumn{1}{l|}{0.09} & \multicolumn{1}{l|}{4.3} & \multicolumn{1}{l|}{0.09} & \multicolumn{1}{l|}{(7,248)} & \multicolumn{1}{l|}{36.87} & \multicolumn{1}{l|}{*\textless{}.001} & 0.51 \\ \hline
\end{tabular}%
\vspace{-0.5em}
%}
\end{table*}

\begin{table*}[htbp]
\centering
\caption{Effect of covariates}
    \vspace{-0.5em}
\label{tab:covariates}
%\resizebox{\textwidth}{!}{%
\begin{tabular}{|l|llrl|llrl|}
\hline
\multirow{2}{*}{} & \multicolumn{4}{l|}{\textbf{Fidelity}} & \multicolumn{4}{l|}{\textbf{Interpretability}} \\ \cline{2-9} 
 & \multicolumn{1}{l|}{\textbf{df}} & \multicolumn{1}{l|}{\textbf{F}} & \multicolumn{1}{l|}{\textbf{Sig.}} & \textbf{eta\textasciicircum{}2} & \multicolumn{1}{l|}{\textbf{df}} & \multicolumn{1}{l|}{\textbf{F}} & \multicolumn{1}{l|}{\textbf{Sig.}} & \textbf{eta\textasciicircum{}2} \\ \hline
BPM expertise & \multicolumn{1}{l|}{(1,248)} & \multicolumn{1}{l|}{0.102} & \multicolumn{1}{r|}{0.749} & 0 & \multicolumn{1}{l|}{(1,248)} & \multicolumn{1}{l|}{0.915} & \multicolumn{1}{r|}{0.34} & 0.004 \\ \hline
Digital literacy & \multicolumn{1}{l|}{(1,248)} & \multicolumn{1}{l|}{0.824} & \multicolumn{1}{r|}{0.365} & 0.003 & \multicolumn{1}{l|}{(1,248)} & \multicolumn{1}{l|}{0.869} & \multicolumn{1}{r|}{0.352} & 0.003 \\ \hline
\textbf{Curiosty} & \multicolumn{1}{l|}{(1,248)} & \multicolumn{1}{l|}{6.175} & \multicolumn{1}{r|}{\textbf{*0.014}} & 0.024 & \multicolumn{1}{l|}{(1,248)} & \multicolumn{1}{l|}{4.691} & \multicolumn{1}{r|}{\textbf{*0.031}} & 0.019 \\ \hline
\textbf{Trust} & \multicolumn{1}{l|}{(1,248)} & \multicolumn{1}{l|}{258.37} & \multicolumn{1}{r|}{\textbf{*0}} & 0.51 & \multicolumn{1}{l|}{(1,248)} & \multicolumn{1}{l|}{233.286} & \multicolumn{1}{r|}{\textbf{*0}} & 0.485 \\ \hline
\end{tabular}%
    \vspace{-1em}
%}
\end{table*}

\section{Predicting citizen's feedback}

While we presented a viable approach for comparing multiple LLMs in the context of generating tax return decisions, we acknowledge that conducting such an empirical evaluation can be tedious and time-consuming. To address this concern, we explored the possibility of taking an additional step to ease such a burden. Specifically, we investigated the feasibility of replacing the empirical effort with a predictive model capable of reliably estimating the average human perceptions of fidelity and interpretability when provided with a textual narrative of an explanation as input.

Our initial intuition for this was ignited by observing the `small multiple' chart depicting distributions of fidelity and interpretability (see Figure~\ref{fig:small-multiple-distributions}) for the 12 explanation versions generated in our experiment. As can be observed, the distributions, though somewhat sparse, exhibit a converging tendency towards a normal curve for each explanation version. This suggests that with a sufficiently large sample, most individuals might agree on fidelity and interpretability scores, with deviations arising from personal characteristics. Consequently, our experimental results also motivated a subsequent effort in which we investigated whether combining the textual explanation with personal traits—such as the background attributes captured in our experiment—could produce a predictive model capable of anticipating human perceptions of its quality.

\subsection{Investigating the Viability of Generating Predictive Models for Fidelity and Interpretability}

We report here an exploration of the feasibility of creating predictive models for average fidelity and interpretability ratings based on a numeric representation of the explanation narratives (embeddings), user characteristics (digital literacy and BPM expertise), and user evaluations. The dataset consisted of user ratings for 12 different versions of explanations about a tax refund process, with each user evaluating two versions. Analyses were conducted using the full dataset and subsets of 90\% and 25\%, alongside various modeling approaches, embedding sources, and dimensionality reduction techniques. Data stratification by different model types was also examined as a means to account for skewed samples.

As a complementary effort, we also explored the possibility of using an LLM model `as-a-judge' to elicit model-based ratings for fidelity and interoperability, having the model prompt include the same explanation and ground truth narratives presented to the human subject in our survey. The LLM mode was instructed to impersonate itself based on the same background characteristics recorded in the survey. For this, we employed GPT-4o and llama-3-1-70b (see prompt in Figure~\ref{lst:prompt-llm-as-a-judge}).

\begin{figure}[ht] % Use H to enforce placement here
\begin{framed}
\begin{MyVerbatim}
\textbf{Evaluate the following text:}
\textbf{Provided text:} "{row['text']}"
\textbf{Based on the Ground truth:} "{row['Ground_truth']}"
\textbf{And by assuming that you are an individual with the below characteristics:}
- Tax payer: {row['is-tax-payer']}
- Gender: {row['gender']}
- Level of education: {row['education']}
- BPM expert: {row['is-bpm-expert']}/7
- Digital skills level: {row['digital-competence']}/7
\textbf{Task:} 
\textbf{Assess the text for the following criteria on a 1-7 Likert scale, where 1 is the lowest and 7 is the highest:}
1. **Fidelity**: How well the provided text aligns with the ground truth.
2. **Interpretability**: How understandable and clear the text is for someone with the specified characteristics.
\textbf{Provide the results as:}
- Fidelity: [score]
- Interpretability: [score]
\end{MyVerbatim}
\vspace{-0.8em}
\end{framed}
\vspace{-1em}
\caption{LLM-as-a-judge for explanations rating}
\vspace{-1em}
\label{lst:prompt-llm-as-a-judge}
\end{figure}

\subsection{Key Results with the Current Dataset}

Linear regression models demonstrated modest explanatory power with the current dataset. For instance, by generating embeddings using term frequency (TF-IDF) within our corpus of 12 explanations and then applying principal component analysis (PCA) to reduce the embedding dimensions to 15, we achieved the following results: For fidelity, the full dataset produced an \( R^2 \) of 0.132, with one embedding variable consistently showing a significant relationship. For interpretability, the model yielded an 
\( R^2 \) of 0.117, with three embedding variables showing modest significance. User characteristics, such as digital competence and BPM expertise, were consistently non-significant in both models.

When using 90\% of the dataset, the fidelity model’s \( R^2 \) increased slightly to 0.150, but the adjusted \( R^2 \) dropped to 0.094, indicating potential overfitting. The interpretability model showed similarly limited performance. On the 25\% sample, \( R^2 \) for fidelity reached 0.273, but adjusted \( R^2 \) dropped to 0.109, with no significant predictors, highlighting instability due to the small sample size. Attempting stratification by LLM model type did not result in significant improvement. Despite a slightly better 
\( R^2 \) in the best-case scenarios, the overall predictive power for both fidelity and interpretability remained very weak (i.e., 0.34 and 0.37, respectively), as reflected by the low mean \( R^2 \) and high RMSE.

\begin{figure}
    \centering
    \includegraphics[width=1\linewidth]{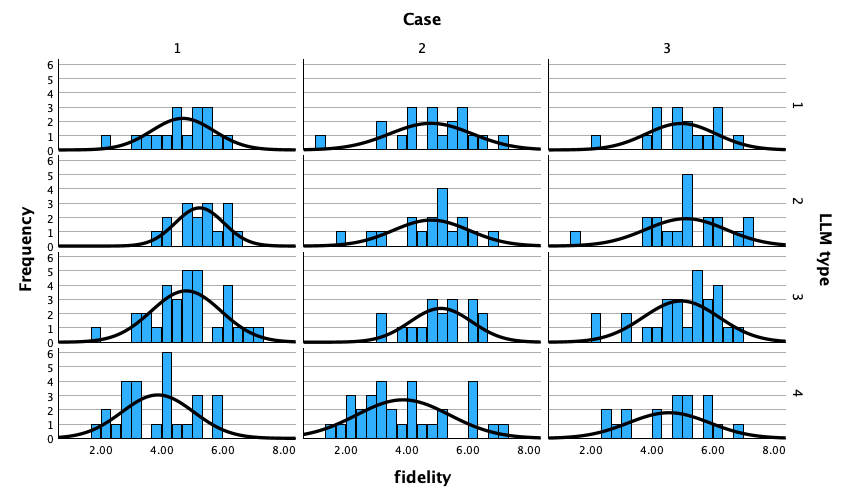}
    \includegraphics[width=1\linewidth]{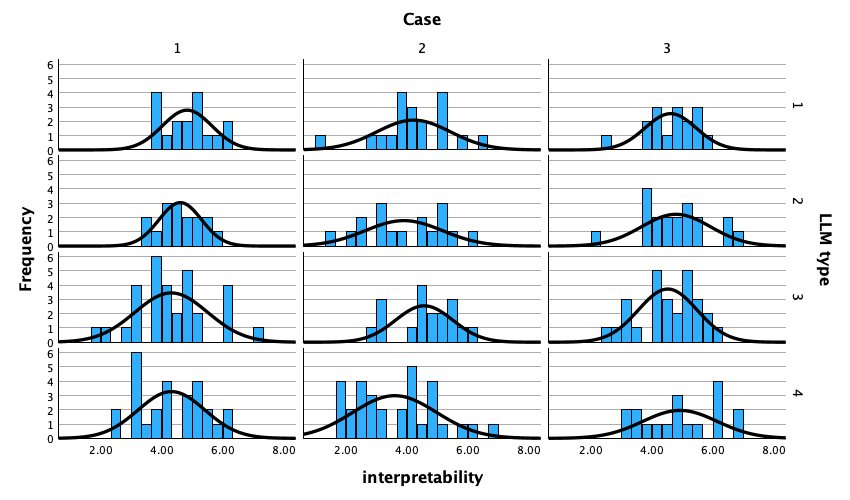}
    \vspace{-1.8em}
    \caption{Distributions for fidelity and interoperability across the different explanation versions also show that data were not yet sufficiently dense to capture the normal distributions that gradually arise from the sample.}
    \vspace{-1em}
    \label{fig:small-multiple-distributions}
\end{figure}

Efforts to improve predictive power included testing embeddings from pre-trained BERT models\footnote{\url{https://huggingface.co/google-bert/bert-base-uncased}} (768 dimensions) and local corpus embeddings (383 dimensions), which were reduced to 5–15 dimensions using PCA. While dimensionality reduction addressed multicollinearity, it did not enhance \( R^2 \) values. Subsequently, we tested several non-linear machine learning models, including Random Forest and XGBoost. However, the results did not enhance \( R^2 \) beyond 0.13, even for non-linear relationships.
As mentioned earlier, sample scarcity is evident in the ``small multiple'' chart in Figure~\ref{fig:small-multiple-distributions}, which displays the data distributions for the various explanation versions generated, with an average of approximately 20 ratings per individual explanation version.

Concerning our effort to use LLMs `as judges' to impersonate human responses, our results showed significant correlations for fidelity using both models (GTP-4o: $r$(254)=.251, p$<$.001, llama-3-1-70b: $r$(254)=.221, p$<$.001), and for interpretability using GPT-4o (GTP-4o: $r$(254)=.132, p=.034).

\subsection{Exploration Recap}

The results underscore the potential of predictive models for fidelity and interpretability but highlight the pressing need for an expanded and more robust dataset. The current limitations, primarily the sparse exposure of users to text versions, constrain the model’s ability to generalize. To achieve a viable model that could stand in place of direct user surveys, a sample size approximately one scale higher, involving around 1,000 users, is strongly recommended. Each user should evaluate a balanced number of text versions (e.g., 4–6) to ensure comprehensive coverage of variations and reduce sampling bias.
The findings suggest that the embeddings already capture valuable predictive signals, as shown by the consistent significance of some of the embedding variables. Additional features, such as linguistic complexity, sentiment, explanation length, and domain-specific metadata, may also aid in achieving better accuracy. With a larger and more diverse sample, it is highly plausible that a generalized model could emerge. Such a model could streamline efforts to assess user perceptions, eliminating the need for repeated surveying while maintaining reliable insights into fidelity and interpretability dimensions.

As per our investigation on using LLMs as judges, our results appear promising. Future work could involve enriched prompt engineering techniques to improve LLM performances, as indicated in the next section.

%\subsection{Model}
%\subsection{Result - prediction accuracy}

\section{Conclusions and Future Work}

This work extends our previous findings on the effectiveness of our developed scale in assessing the perceived quality of explanations for process outcomes and conditions. We demonstrate another valuable application of the scale to facilitate the selection of an effective LLM, thereby further promoting its instrumental validity while remaining consistent with the original identification of trust, curiosity, and moderating factors. Additionally, as a concrete exemplar, we show the applicability of the scale to the e-government case of the tax refund process. Similarly, the survey could be adapted to other e-government services. For replicability, the survey forms, prompts, and collected data are accessible at~\cite{Survey2025}.

While using our developed survey to facilitate the selection among multiple LLMs provides ordinal insights into perceived fidelity and interoperability, the final choice ultimately depends on the weight preferences assigned to these two dimensions.
Eliciting such weighting may not be straightforward for IT infrastructure designers. However, conventional multi-objective decision analysis techniques could likely be employed as a final step, also incorporating other operational factors such as cost and computational resources. 

Concerning our exploration of creating a predictive model, future work could focus on scaling up data collection, fine-tuning embeddings on task-specific corpora to align more closely with the constructs of fidelity and interpretability, extended feature engineering, and employing mixed-effects models to better account for variability between users and texts. One way to achieve scalability is by fostering an ecosystem for the exchange of survey results among communities of interest that focus on similar target populations. This type of data aggregation could be enhanced by integrating a survey platform like ECSurvey with a data exchange platform such as FAME (an outcome of the FAME EU project\footnote{\url{https://www.fame-horizon.eu/the-project/}}). This approach can particularly benefit small to mid-sized organizations that may lack the resources to conduct a full-scale user study.
With these improvements, the vision of a robust, generalized predictive model appears not only promising but achievable, making it a worthwhile investment for organizations aiming to reduce the effort and cost of user surveys.

Concerning our exploration of using LLMs as judges, future improvements could include having the LLM provide more detailed scores for the dimensions underlying fidelity and interpretability, as outlined in the survey. Additional enhancements might involve few-shot training with real survey responses and refining the prompt to request a numeric rating on a continuous 1-7 scale rather than discrete values. Furthermore, leveraging domain-specific knowledge by incorporating contextual documents (via RAG) related to the process could enable LLMs to gain a deeper understanding of the domain.

% \begin{table}[htbp]
% \caption{Table Type Styles}
% \begin{center}
% \begin{tabular}{|c|c|c|c|}
% \hline
% \textbf{Table}&\multicolumn{3}{|c|}{\textbf{Table Column Head}} \\
% \cline{2-4} 
% \textbf{Head} & \textbf{\textit{Table column subhead}}& \textbf{\textit{Subhead}}& \textbf{\textit{Subhead}} \\
% \hline
% copy& More table copy$^{\mathrm{a}}$& &  \\
% \hline
% \multicolumn{4}{l}{$^{\mathrm{a}}$Sample of a Table footnote.}
% \end{tabular}
% \label{tab1}
% \end{center}
% \end{table}

%\section*{Acknowledgment}

\bibliographystyle{IEEEtran}  
\bibliography{references}

\begin{thebibliography}{10}
\providecommand{\url}[1]{#1}
\csname url@samestyle\endcsname
\providecommand{\newblock}{\relax}
\providecommand{\bibinfo}[2]{#2}
\providecommand{\BIBentrySTDinterwordspacing}{\spaceskip=0pt\relax}
\providecommand{\BIBentryALTinterwordstretchfactor}{4}
\providecommand{\BIBentryALTinterwordspacing}{\spaceskip=\fontdimen2\font plus
\BIBentryALTinterwordstretchfactor\fontdimen3\font minus \fontdimen4\font\relax}
\providecommand{\BIBforeignlanguage}[2]{{%
\expandafter\ifx\csname l@#1\endcsname\relax
\typeout{** WARNING: IEEEtran.bst: No hyphenation pattern has been}%
\typeout{** loaded for the language `#1'. Using the pattern for}%
\typeout{** the default language instead.}%
\else
\language=\csname l@#1\endcsname
\fi
#2}}
\providecommand{\BIBdecl}{\relax}
\BIBdecl

\bibitem{Alminshid2021FactorsSector}
K.~Alminshid and M.~Omar, ``{Factors affecting employees' adoption of e-government in the Iraqi public education sector},'' \emph{Electronic Government}, vol.~17, no.~2, 2021.

\bibitem{Dashti2010TrustPerspective}
A.~Dashti, I.~Benbasat, and A.~Burton-jones, ``{Trust , Felt Trust and E-Government Adoption : A Theoretical Perspective},'' \emph{Working Papers on Information Systems}, vol.~10, no.~83, 2010.

\bibitem{Carter2005TheFactors}
L.~Carter and F.~B{\'{e}}langer, ``{The utilization of e-government services: Citizen trust, innovation and acceptance factors},'' \emph{Information Systems Journal}, vol.~15, no.~1, 2005.

\bibitem{Dumas2023}
M.~Dumas, F.~Fournier, L.~Limonad, A.~Marrella, and {et al.}, ``{AI-augmented Business Process Management Systems: A Research Manifesto},'' \emph{ACM Transactions on Management Information Systems}, vol.~14, no.~1, 2023.

\bibitem{Afroogh2024}
S.~Afroogh, A.~Akbari, E.~Malone, M.~Kargar, and H.~Alambeigi, ``{Trust in AI: progress, challenges, and future directions},'' \emph{Humanities and Social Sciences Communications}, vol.~11, no.~1, p. 1568, 11 2024.

\bibitem{Abdul2024}
N.~S. Abdul~Wahi and L.~Berenyi, ``{Examining the Effect of Social Influence and Facilitating Conditions on E-government Adoption by Employees in Mandatory Condition},'' in \emph{Proceedings of the Central and Eastern European eDem and eGov Days 2024}.\hskip 1em plus 0.5em minus 0.4em\relax New York, NY, USA: ACM, 9 2024, pp. 104--110.

\bibitem{Fahland2024}
\BIBentryALTinterwordspacing
D.~Fahland, F.~Fournier, L.~Limonad, I.~Skarbovsky, and A.~J.~E. Swevels, ``{How well can large language models explain business processes as perceived by users?}'' \emph{Data {\&} Knowledge Engineering}, vol. 157, p. 102416, 2 2025. [Online]. Available: \url{https://www.sciencedirect.com/science/article/pii/S0169023X25000114}
\BIBentrySTDinterwordspacing

\bibitem{Fournier2023v3}
\BIBentryALTinterwordspacing
F.~Fournier, L.~Limonad, I.~Skarbovsky, and Y.~David, ``{The WHY in Business Processes: Discovery of Causal Execution Dependencies},'' \emph{K{\"{u}}nstliche Intelligenz}, 1 2025. [Online]. Available: \url{https://rdcu.be/d52Qz}
\BIBentrySTDinterwordspacing

\bibitem{Weske2019}
\BIBentryALTinterwordspacing
M.~Weske, ``{Business Process Management Architectures},'' in \emph{Business Process Management}.\hskip 1em plus 0.5em minus 0.4em\relax Berlin, Heidelberg: Springer Berlin Heidelberg, 2019, pp. 351--384. [Online]. Available: \url{http://link.springer.com/10.1007/978-3-662-59432-2\_8}
\BIBentrySTDinterwordspacing

\bibitem{vanderAalst2016ProcessMining}
\BIBentryALTinterwordspacing
W.~van~der Aalst, \emph{{Process Mining}}.\hskip 1em plus 0.5em minus 0.4em\relax Berlin, Heidelberg: Springer, 2016. [Online]. Available: \url{http://link.springer.com/10.1007/978-3-662-49851-4}
\BIBentrySTDinterwordspacing

\bibitem{Adadi2018}
A.~Adadi and M.~Berrada, ``{Peeking Inside the Black-Box: A Survey on Explainable Artificial Intelligence (XAI)},'' \emph{IEEE Access}, vol.~6, 2018.

\bibitem{Meske2022}
C.~Meske, E.~Bunde, J.~Schneider, and M.~Gersch, ``{Explainable Artificial Intelligence: Objectives, Stakeholders, and Future Research Opportunities},'' \emph{Information Systems Management}, vol.~39, no.~1, pp. 53--63, 1 2022.

\bibitem{Rehse2019}
J.~R. Rehse, N.~Mehdiyev, and P.~Fettke, ``{Towards Explainable Process Predictions for Industry 4.0 in the DFKI-Smart-Lego-Factory},'' \emph{KI - Kunstliche Intelligenz}, vol.~33, no.~2, 2019.

\bibitem{Guidotti2018}
R.~Guidotti, A.~Monreale, S.~Ruggieri, F.~Turini, F.~Giannotti, and D.~Pedreschi, ``{A survey of methods for explaining black box models},'' \emph{ACM Computing Surveys}, vol.~51, no.~5, 2018.

\bibitem{GartnerGenAI}
\BIBentryALTinterwordspacing
``{Definition of Generative AI - Gartner Information Technology Glossary}.'' [Online]. Available: \url{https://www.gartner.com/en/information-technology/glossary/generative-ai}
\BIBentrySTDinterwordspacing

\bibitem{Gartner2024}
A.~Chandrasekaran and L.~Ramos, ``{Hype Cycle for Generative AI (ID G00812271)},'' Gartner, Tech. Rep., 7 2024.

\bibitem{GartnerLLM}
\BIBentryALTinterwordspacing
``{Definition of Large Language Models (LLMs) - Gartner Information Technology Glossary}.'' [Online]. Available: \url{https://www.gartner.com/en/information-technology/glossary/large-language-models-llm}
\BIBentrySTDinterwordspacing

\bibitem{LLMprompt2023}
P.~Liu, W.~Yuan, J.~Fu, Z.~Jiang, H.~Hayashi, and G.~Neubig, ``{Pre-train, Prompt, and Predict: A Systematic Survey of Prompting Methods in Natural Language Processing},'' \emph{ACM Computing Surveys}, vol.~55, no.~9, 2023.

\bibitem{Survey2025}
\BIBentryALTinterwordspacing
L.~Limonad and F.~Fournier, ``{Explanation Quality Survey - the Tax Refund Case},'' 1 2025. [Online]. Available: \url{https://zenodo.org/records/14637610}
\BIBentrySTDinterwordspacing

\end{thebibliography}

\end{document}